\documentclass[showpacs,superscriptaddress, preprintnumbers,amsmath,amssymb]{revtex4}

\pdfoutput=1

\usepackage{graphicx}
\usepackage{dcolumn}
\usepackage{bm}
\include{epsf}

\begin{document}


\title{A model of ideological struggle}

\author{Nikolay K. Vitanov}\email{vitanov@imbm.bas.bg}
\affiliation{Institute of Mechanics, Bulgarian Academy of Sciences,
Akad. G. Bonchev Str., Bl. 4, 1113 Sofia, Bulgaria
}

\author{Zlatinka I. Dimitrova}
\affiliation{Institute of Solid State Physics, Bulgarian Academy of Sciences,
Blvd. Tzarigradsko Chaussee 72, 1784, Sofia, Bulgaria}
 
\author{Marcel Ausloos}\email{marcel.ausloos@ulg.ac.be}
\affiliation{GRAPES, B5 Sart-Tilman, B-4000 Li\`ege, Euroland
}


\begin{abstract}
A general model for opinion formation and competition, like in ideological struggles is 
formulated. The underlying set is a closed one, like a country but in which the population 
size is variable in time. Several   ideologies  compete to increase their number of adepts.
Such followers can be either converted from one ideology to another or become followers of an 
ideology though being previously ideologically-free. A reverse process is also allowed. 
We consider two kinds of conversion: unitary  conversion, e.g. by means of mass 
communication tools, or binary conversion, e.g. by means of interactions between people.
It is found that  the steady state,when it exists, depends on the number of ideologies. 
Moreover when the number of ideologies increases some tension arises between
them. This tension can change in the course of time. 
We propose to measure the ideology tensions through an appropriately defined scale index. 
\noindent

\end{abstract}



\maketitle

\section{Introduction}
\label{intro}
\subsection{Nonlinear systems, opinion formation and population dynamics}
The case of opinion formation \cite{gal07} - 
\cite{holyst01}
in sociology is seen in recent physics works
as analogous to state phase evolution in non-equilibrium systems.
Applications to population dynamics \cite{hassel}, extinction of populations \cite{pec08},
animal and human migration \cite{vit09}, policy and politics \cite{moss}, 
languages \cite{lang1},
religions \cite{petr07} - 
\cite{hayw05} 
are found to be similar to epidemics \cite{anderson},
forest fires  \cite{Duarte} and other self-organizing systems much studied in 
statistical physics.
No general pattern is however available and situations can be very varied. Whence
it seems appropriate to continue considering the questions through the methods used in the
theory of turbulence \cite{berge} - 
\cite{sagdeev}, 
low-dimensional dynamical systems  \cite{novak} - 
\cite{ott} 
or theory of nonlinear waves \cite{dodd} -  
\cite{remos} 
but {\it exordium} to go back to the classical  Verhulst ideas and Lotka-Volterra model 
by introducing some
realistic conditions on the growth ratios and on the interaction coefficients  between  the 
populations \cite{dv00} - 
\cite{dv06}.
Indeed there is then a connexion to the problem of extinction of populations \cite{pec08},
religions \cite{petr07}, languages \cite{abrams}, and to the very modern  question of
internet governance in which the {\it old} stakeholders, i.e. the most powerful actors,
and  a variable set of {\it new} participants are somewhat abused or lacking cohesion in their reaction
\cite{olivera}.
\par
Interestingly this demands at some stage a consideration of the connexion between
economical and social physics and social dynamics, including the analysis of time series
\cite{aruka06}  - 
\cite{fab05}
though with some caveat due to either size or debatable data value. Whence the need for a theoretical approach
and some modelization in order to focus any data gathering toward useful input in further work.
\subsection{Organization of the paper}
In Sect. 2, we formulate a general/mathematical model
in a finite system size, {\it we emphasize,}  allowing for a changing population size. In a realistic way we
consider people unaffected by the available ideologies as well as conversions to ideologies. 
Two mechanisms are discussed by which the followers of an ideology can increase:
{\it unitary} conversion (a citizen is converted  by means of forms of mass
 communication such as newspapers, radio or television channels) and {\it
binary} conversion  (a citizen is converted by interpersonal contacts with 
other citizens). 
\par
In Sect. 3 we study the case when only one ideology spreads among the population. 
The whole  country population is found to evolve  towards an  equilibrium state.
In this state some part of the people becomes followers of the ideology and the remaining ones
are not followers of the ideology. The fraction of  adepts depends on the intensity
of the unitary and binary conversion as well as on the ability of the ideology to
reduce the dissatisfaction among its followers.  
In Sect. 4, the case of two ideologies 
is examined. The introduction of a second ideology leads to tensions between the 
ideologies as the number of followers drops in comparison to the case when each of the
ideologies is alone in the country. The tension can be quantified on a scale that 
can be considered  to be an index of ideological tensions in the society. 
\par
In Sect. 5, in the case of three ideologies,  we  briefly show that a chaotic behavior 
is seen to occur among many other solutions. This seems close to intuition. 
In Sect. 6, conclusions are outlined and connected for relevance sake to a usual observation 
that the coexistence of ideologies implies the existence of tensions between them. 

\section{Mathematical formulation of the model}
\label{sec:2}
Let  us consider a set ( country ) with a population of $N$ agents.
We are going to consider that the population   is divided into $n+1$ factions: $n$ factions each with a 
different specific ideology,  such that  the
number of members in the corresponding populations  are $N_{1}, N_{2}, \dots, N_{n}$, and a
fraction $N_{0}$ of people which are not followers of  any ideology at a given 
moment of time. Then
\begin{equation}
N = N_{0} + \sum_{i=1}^{n} N_{i}
\end{equation} 
We assume that the overall population
evolves according to the generalized Verhulst law
\begin{eqnarray}
\frac{dN}{dt}=r(t,N,N_{1}, \dots, N_{n},p_{\mu}) N \times \nonumber\\
 \left[ 1 - \frac{N}{C(t,N,N_{1},
\dots, N_{n},p_{\mu})} \right] 
\end{eqnarray}
where $p_{\mu}=(p_{1},\dots,p_{m})$ are parameters, describing the environment. 
\par 
The growth process is constantly disrupted by small extinction events, as in \cite{wilke}, 
monitored  through $r(t, N, N_{1},$ 
$\dots, N_{n}, p_{\mu})$. $r$ is the overall population growth rate; 
$r$ can be positive or negative. In this paper we shall
consider the case $r>0$, i.e. we shall study the spreading and competition between ideologies
in a country with a growing total population. $C(t,N,N_{1},\dots, N_{n},p_{\mu})$
 is the maximum possible population of the country (its so called carrying capacity). 
In every ideological population $i$ we have to account for the  following processes:
deaths, dissatisfaction, unitary conversion, and binary conversion.
\begin{enumerate}
\item
First, we expect a
decrease of the number of  followers of   an  ideology through death or  dissatisfaction
with the ideology,  i.e. through a term $r_{i} N_{i}$, where $r_{i} \le 0$. 
In general $r_{i}=r_{i}(t,N,N_{1},\dots, N_{n},p_{\mu})$.
\item
Unitary conversion: such a conversion from one ideology to another 
is made without direct contact between the followers of different ideologies.
The conversion happens through the information environment of the population.
Elements of this environment for  example are the newspapers, the radio stations,  
television channels, printed propaganda materials or mass events such as speeches during
election campaigns. Excluded are the direct interpersonal contacts which
lead to the binary conversion described below. In order to model the unitary
conversion we assume that the number of people converted from ideology $j$ to ideology $i$
is proportional to the number $N_{j}$ of the followers of the ideology $j$.  An 
$f_{ij}$  coefficient characterizes the intensity  with which this conversion occurs.
The corresponding modeling term is $f_{ij} N_{j}$. We assume that $f_{ii}=0$.
In addition a term $f_{i0} N_{0}$   describes the unitary conversion  
toward ideology $i$ from the $N_{0}$ people who were not followers of any ideology 
at the corresponding moment of time. In general
$$
f_{ij}=f_{ij}(t,N,N_{1},\dots,N_{n},p_{\mu},C)
$$ 
$$
f_{i0}=f_{i0}(t,N,N_{1},\dots,N_{n},p_{\mu},C)
$$
\item
Binary conversion: $ b_{ijk} N_{j} N_{k}$. In general this term
describes the conversion to the
$i$-th ideology because of direct interaction 
between members of the $j$-th and $k$-th ideology. 
We assume that the intensity of the interpersonal contacts is
proportional to the numbers $N_{j}$ and $N_{k}$ of the followers of the two ideologies. 
The coefficient characterizing the intensity of the binary conversion  is  $b_{ijk}$.
The larger is $b_{ijk}$, the more people are converted to the $i$-th ideology. 
In   general  the binary conversion coefficients can be
$b_{ijk}=b_{ijk}(t,N,N_{1},$ $ \dots, N_{n}, p_{\mu},C)$. 
Of course $b_{iii}=0$: there is no self-conversion.
\end{enumerate}
The equation for the evolution of the followers of the ideology $i$ becomes
\begin{eqnarray}\label{change_followers}
\frac{dN_{i}}{dt} = 
r_{i}(t,N,N_{1},\dots,N_{n},p_{\mu},C) N_{i} 
 + \nonumber\\ f_{i0}(t,N,N_{1},\dots,N_{n},p_{\mu},C) N_{0} + 
 \nonumber\\
\sum_{j=1}^{n} f_{ij} (t,N,N_{1},\dots,N_{n},p_{\mu},C) N_{j} + \nonumber\\
b_{i0}(t,N,N_{1},\dots,N_{n},p_{\mu},C) N_{i} N_{0} + 
\nonumber\\
\sum_{j=1}^{n} \sum_{k=1}^{n} b_{ijk}(t,N,N_{1},\dots,N_{n},p_{\mu},C) N_{j} N_{k} 
\end{eqnarray}
In general we can have a co-evolution of the environment and the populations, i.e.
$p_{\mu} = p_{\mu}(N,N_{1},\dots,N_{n},C,t)$, but this  will not be discussed here.
\par
Indeed in this paper we shall discuss the simplest version of the model namely
the case in which all the coefficients 
 are time  and $p_{\mu}$ independent.
Then the model system becomes geared by
\begin{equation}\label{mod_syst1}
N=N_{0}+ \sum_{i=1}^{n} N_{i} 
\end{equation}
\begin{equation}\label{mod_syst2}
\frac{dN}{dt} = rN \left(1- \frac{N}{C} \right)
\end{equation}
\begin{equation}\label{mod_syst3}
\frac{dN_{i}}{dt} = r_{i} N_{i} + f_{i0} N_{0}+ \sum_{j=1}^{n} f_{ij} N_{j}+
b_{i0} N_{i} N_{0} + \sum_{j=1}^{n} \sum_{k=1}^{n} b_{ijk} N_{j} N_{k} 
\end{equation}
\par
Two  remarks are in order. 
\begin{enumerate}
\item
Notice that arbitrary values are not allowed for the coefficients of the model. They must 
have   values  such that  $N$, $N_{0}$, $N_{1}$, $\dots$, $N_{n}$  
be nonnegative at each $t$.
\item
Let us consider the $i$-th population and the binary
conversion characterized by the coefficients $b_{ijk}$ and $b_{ikj}$ where $j$ and $k$
are different from $i$. One could at first think that $b_{ijk} N_{j}N_{k}$
and $b_{ikj} N_{k} N_{j}$ describe one and the same 
process in which the interaction between followers of the $j$-th and $k$-th ideology
leads to a conversion of these to the $i$-th ideology. In general  however one should not 
identify the two terms. In so doing in the general model 
we retain one additional degree of freedom,
i.e., that which allows to distinguish between the ideology that is of the initiator of the
interaction and the ideology of someone who is apparently simply a participant  
in the interaction.
\end{enumerate}
\par
Below we shall consider the dynamics of populations of followers of the ideologies for the
cases of presence of 1,2 or 3 ideologies in the country.
\section{The case of one ideology}
In the case of spreading of one ideology the population of the country  is 
divided into two groups: $N_{1}$ followers of the single ideology   
and $N_{0}$ people who are not followers of this ideology at the corresponding  time.
Let us first discuss the case when only the unitary conversion scheme exists,  as possibly 
moving the  ideology-free population toward the single ideology, i.e. $f_{10}$ is finite.
Let the initial conditions be  $N(t=0)=N(0)$
and $N_{1}(t=0)=N_{1}(0)$. The solution of the model system is
\begin{equation}\label{solution1}
N(t) = \frac{C N(0)}{N(0) + (C-N(0))e^{-rt}}
\end{equation}
\begin{eqnarray}\label{solution2}
N_{1}(t) = e^{-(f_{10}-r_{1})t} \bigg \{ N_{1}(0) + \frac{c f_{10}}{r} \times \nonumber\\
\bigg [ 
{\Phi}\bigg ( -\frac{C-N(0)}{N(0)},1,-\frac{f_{10}-r_{1}}{r} \bigg ) -
\nonumber \\
e^{t(f_{10}-r_{1})} {\Phi} \bigg(-\frac{C-N(0)}{N(0) e^{rt}},1,
- \frac{f_{10}-r_{1}}{r}\bigg) \bigg] \bigg\}
\end{eqnarray}
\begin{equation}\label{solution3}
N_{0}(t)=N(t)-N_{1}(t)
\end{equation}
where $\Phi$ is the special function
\begin{equation}\label{lerh_phi}
{\Phi} (z,a,v) = \sum_{n=0}^{\infty} \frac{z^{n}}{(v+n)^{a}} \hskip.1cm ; \hskip.5cm 
\mid z \mid < 1
\end{equation} 
The  condition $\mid z \mid < 1 $ is equivalent to $N^{*}>C/2$.
This means that the solution (7) describes the development in  the case of  a densely
populated territory. For $N^{*} \le C/2$, $\Phi$ can tend to $\infty$ and  
we can only obtain a  numerical solution of the model system of equations.
\par
The obtained solution describes an evolution in which the total population $N$ reaches
 asymptotically the carrying capacity $C$ of the environment. The number of adepts of
the ideology reaches an equilibrium value which
corresponds to the fixed point of the model equation for $\frac{d N_{1}}{dt}$.
This fixed point is 
$$
\hat{N}_{1} = \frac{C f_{10}}{f_{10} -r_{1}}
$$
The number of people which are not followers of the ideology asymptotically
tends to  $N_{0}=C-\hat{N}_{1}$.
As a numerical example let  $C=1$, $f_{10}=0.03$ and $r_{1}=-0.02$, 
then $\hat{N}_{1}=0.6$ which
means that the evolution of the system leads to an asymptotic state in which
60 \% of the population are followers of the ideology and 40 \%  are not.  
\par
Now let  not only unitary but also binary conversion processes be possible.
The evolution in this case cannot be investigated analytically. However the asymptotic
 behavior for  $N_{1}$ can be obtained when the total population $N$ has reached the
carrying capacity $C$ of the environment. For this asymptotic state
the evolution of $N_{1}$ reads
\begin{equation}\label{single_full}
\frac{d N_{1}}{d t} = r_{1} N_{1} + f_{10} (C - N_{1}) + b_{10} N_{1} (C-N_{1})
\end{equation}
There exist two fixed points but only one of them satisfies the requirement
$\hat{N}_{1}>0$. This fixed point is 
\begin{equation}\label{full_fp}
\hat{N}_{1}= \frac{(r_{1}-f_{10}+b_{10}C) + \sqrt{(r_{1}-f_{10}+b_{10}C)^{2}+ 
4b_{10} f_{10}C}}{2 b_{10}}
\end{equation}
The equation (\ref{single_full}) has an analytical solution.
The 
solution depends on whether  $N_{1}>\hat{N}_{1}$ or $N_{1}<\hat{N}_{1}$.
The two cases can be realized respectively when $N_{1} (0) > \hat{N}_{1}$ and
$N_{1}(0)<\hat{N}_{1}$. If $N_{1}(0)>\hat{N}_{1}$ then
\begin{equation}\label{full1}
N_{1}= \frac{X_{1}}{Y_{1}}
\end{equation}
where
\begin{eqnarray}
X_{1}=r_{1}-f_{10}+b_{10}C +
\sqrt{(r_{1}-f_{10}+b_{10}C)^{2}+ 4 b_{10} 
f_{10}C}+ \nonumber \\
e^{-(t+ \tau )(r_{1}-f_{10}+b_{10}C)} \times \nonumber\\
\bigg( r_{1}-f_{10}+b_{10}C -
\sqrt{(r_{1}-f_{10}+b_{10}C)^{2} + 4 b_{10} f_{10}C} \bigg) \nonumber
\end{eqnarray}
\begin{eqnarray}
Y_{1}=2 b_{10}
\bigg(1 - e^{-(t+\tau )} (r_{1}-f_{10}+b_{10}C) \bigg)
\nonumber
\end{eqnarray}
and where the characteristic time $\tau$ is here given by
\begin{eqnarray}
\tau = - \frac{1}{r_{1}-f_{10}+b_{10}C} \times \nonumber\\
\ln \bigg( \frac{2 b_{10} N_{1}(0)+ Z}{2 b_{10} N_{1}(0) +
r_{1} - f_{10} + b_{10}C - Z} \bigg)
\nonumber \\
\end{eqnarray}
where $Z=\sqrt{(r_{1}-f_{10}+b_{10}C)^{2}+4b_{10} f_{10}C}$.
\par
For the case $0<N_{1}(0) <\hat{N}_{1}$
\begin{eqnarray}
N_{1}=\frac{X_{2}}{Y_{2}}
\end{eqnarray}
where
\begin{eqnarray}
X_{2}= X_{1} \nonumber
\end{eqnarray}
\begin{eqnarray}
Y_{2}=2 b_{10}
\bigg(1 + e^{-(t+\tau )} (r_{1}-f_{10}+b_{10}C) \bigg) \nonumber
\end{eqnarray}

\begin{figure}
\centering
\includegraphics[height=5cm,width=5cm]{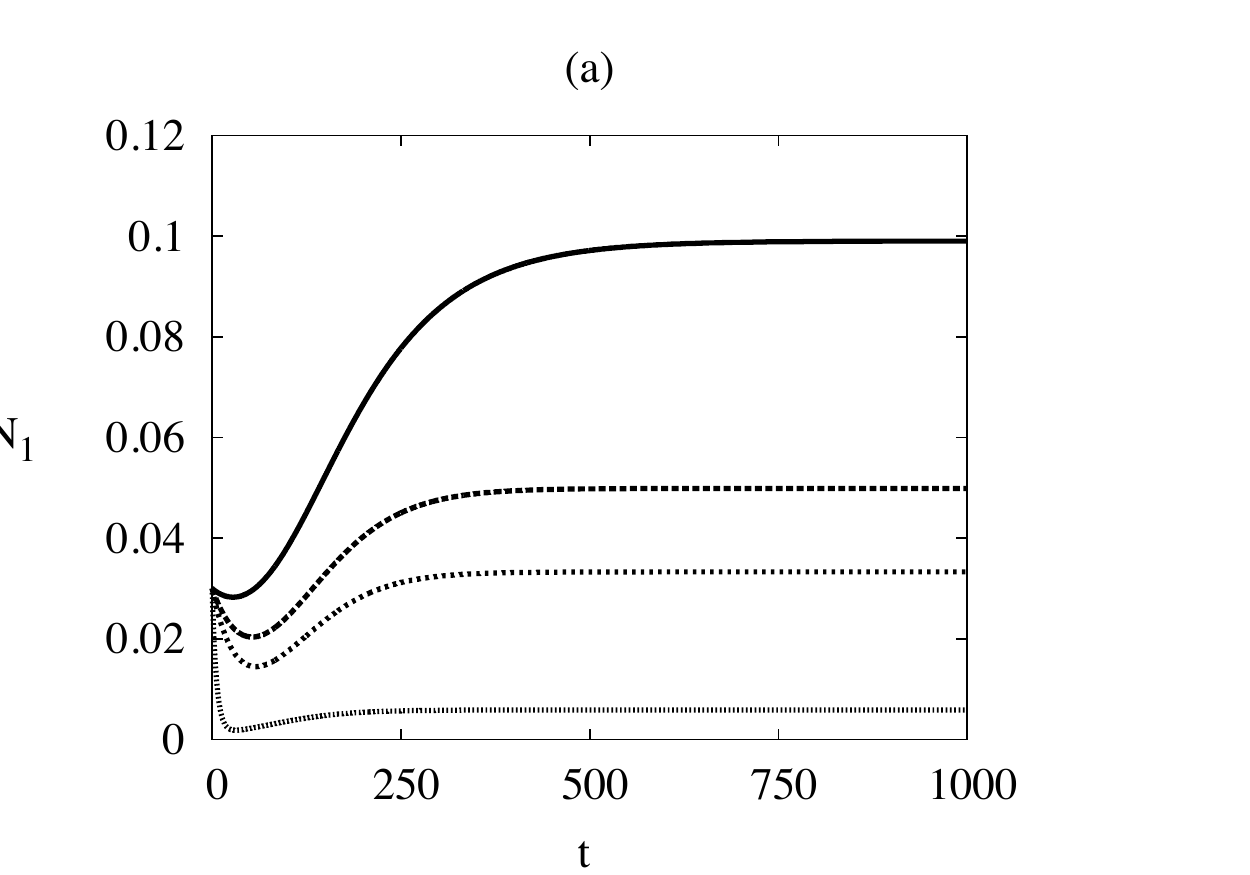}
\includegraphics[height=5cm,width=5cm]{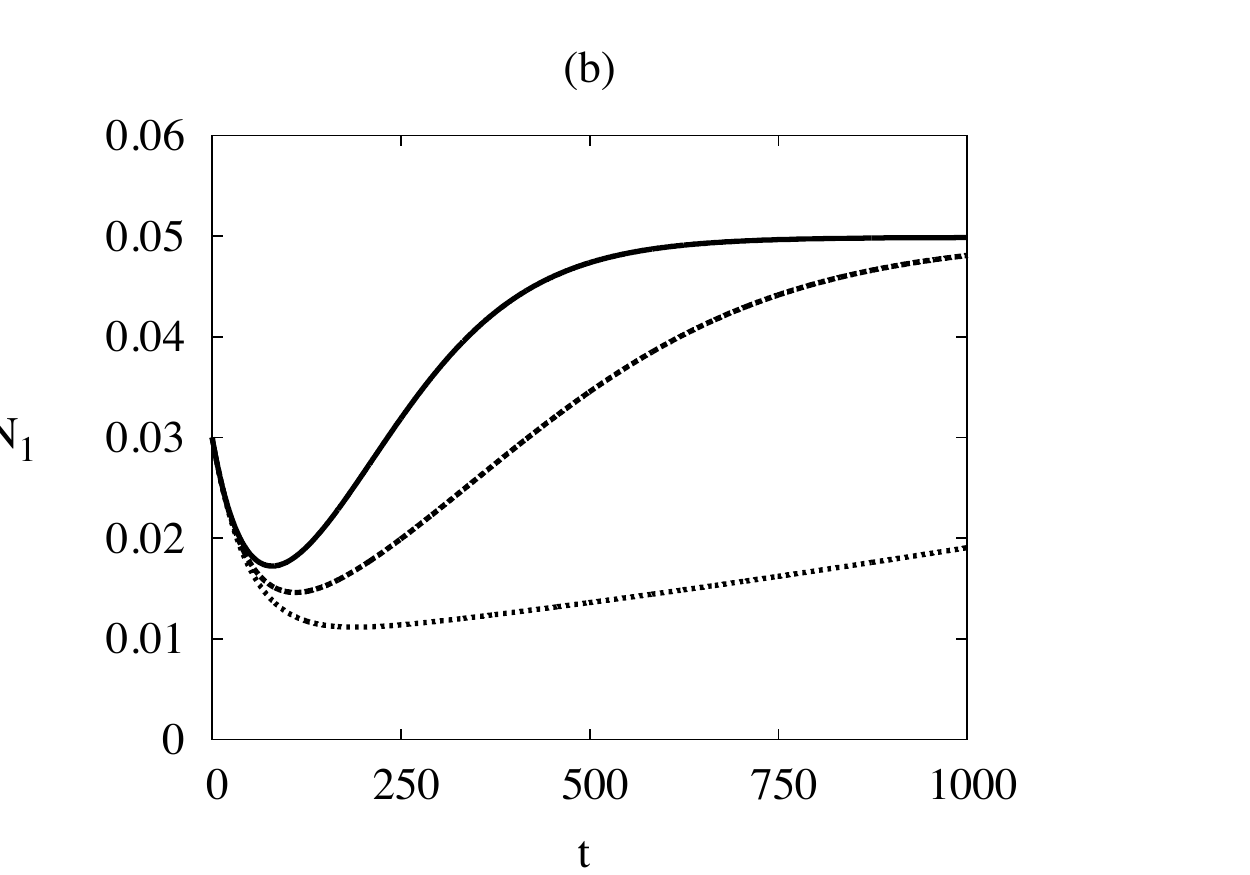}
\caption{\label{fig1.2.dec} (a): Illustration of the inertial growth  and its dependence on the parameter
$r_{1}$. The other parameters are: $C=1$, $r=0.2$, $f_{10}=b_{10}=0.001$.
(b):  Illustration of the inertial growth  and its dependence on the parameter
$r$, with $C=1$, $r_{1}=-0.02$, $f_{10}=b_{10}=0.01$}
\end{figure}

\par
Let us now discuss the time behavior of the number of  followers of the
ideology. There are three possibilities
\begin{enumerate}
\item
$\frac{d N_{1}}{dt} >0$ for all $t$, i.e. the number of followers 
increases monotonically. For the
particular case of  only a unitary conversion process the condition reads
$-\frac{r_{1}}{f_{10}} < \frac{N_{0}}{N_{1}}$
\item
$\frac{dN_{2}}{dt} <0$ for all $t$, i.e. the number of followers decreases monotonically.
For the case  of only  unitary conversions, the condition reads
$-\frac{r_{1}}{f_{10}} > \frac{N_{0}}{N_{1}}$
\item
The most interesting case is when $\frac{d N_{1}}{dt}$ can change  sign with
increasing $t$. The following effect can be observed: the number of followers of the
ideology can increase despite the fact that $r_{1}<0$. The reason for this effect
is the increasing number $N_{0}$, occurring  because of the fast
enough growth of the population of the country. When $N_{0}$ is small the term containing
$r_{1}$ dominates and $N_{1}$ decreases. But in the course of time $N_{0}$ increases.
Then the conversion begins to dominate over dissatisfaction and the number of the
followers of the ideology begins to increase.
We shall call such a kind of  growth of the
followers of the ideology an inertial growth.  
\end{enumerate}
Fig. 1 illustrates an inertial growth. In Fig. 1a,  it can be observed that the  process of
initial shrinking and then of inertial growth can 
exist for  a large 
range of  coefficient values. 
Inertial growth can exist even if the ideology is weak
with respect to its keeping of followers, i.e.
when $r_{1}$ has large negative values.
Note that the figure illustrates the case when the 
carrying capacity of the environment is a constant. If the carrying capacity  would change, 
one could observe sequences of phases of inertial growth and shrinking.
Fig. 1b  shows that  a small growth rate $r$, i.e. of the total population,
leads to a slowing down of  the inertial growth process. 
%
\section{Case of two ideologies: The ideological tension}
In this section we will discuss the competition for adepts that the presence of a second 
ideology introduces; this is leading to  a measurable conflictual tension, as will be shown.
\par
Let us consider the model system for the case of two ideologies with populations of
followers $N_{1}$ and $N_{2}$; we assume that all parameters are kept constant. We have
\begin{equation}\label{mod1}
\frac{dN}{dt} = r N \left( 1- \frac{N}{C} \right)
\end{equation}
\begin{eqnarray}\label{mod2}
\frac{dN_{1}}{dt}= r_{1} N_{1} + f_{10} N_{0} + f_{12} N_{2} + b_{10}N_{0} N_{1} +
 \nonumber\\
(b_{112}+b_{121}) N_{1} N_{2} + b_{122} N_{2}^{2}
\end{eqnarray}
\begin{eqnarray}\label{mod3}
\frac{dN_{2}}{dt}=r_{2} N_{2} + f_{20} N_{0} + f_{21} N_{1} + b_{20} N_{0} N_{2} + 
\nonumber\\
b_{211}
 N_{1}^{2} + (b_{212}+b_{221}) N_{1} N_{2}
\end{eqnarray}
\begin{equation}\label{mod4}
N= N_{0}+ N_{1} + N_{2}
\end{equation}
Let us quantify the tension between  ideologies by means of  an asymptotic analysis. 
We discuss the case  of  only  
unitary conversion of members of the population that are not followers 
of any of the ideologies. In order to emphasize the unitary conversion effects   let us 
assume that the binary conversion  as well as the unitary conversion 
from one ideology to the the other one are negligible. 
In this case $f_{12}=f_{21}=0$ and
 $b_{112}=b_{122}=b_{121}=b_{211}=b_{212}=b_{221}=0$. In addition  let us  consider the
asymptotic case in which the total population has reached the carrying capacity :
$N=C$. The   equilibrium state is characterized by the fixed points :
\begin{equation}\label{fixpoints}
\breve{N}_{1}=\frac{C r_{2} f_{10}}{r_{1} f_{20} + f_{1} r_{2} - r_{1} r_{2}}, \hskip.25cm
\breve{N}_{2}= \frac{C r_{1} f_{20}}{r_{1} f_{20} + f_{1} r_{2} - r_{1} r_{2}}
\end{equation}
If the ideologies were without competition their size would be; see the previous section
\begin{equation}\label{fixpoints1}
\hat{N}_{1} = \frac{C f_{10}}{f_{10}-r_{1}}; \hskip.25cm 
\hat{N}_{2} = \frac{C f_{20}}{f_{20}-r_{2}}
\end{equation}
It can thus be observed  that the popularity of an ideology shrinks when
competing ideology or ideologies spread around the country. Let us evaluate this
shrinking. We have
\begin{equation}\label{shrinking}
 \frac{\breve{N}_{1}}{\hat{N_{1}}} = \frac{1}{1+\frac{r_{1} f_{20}}{r_{2}(f_{10}- r_{1})}}, 
\hskip.25cm
 \frac{\breve{N}_{2}}{\hat{N_{2}}} = \frac{1}{1+\frac{r_{2} f_{10}}{r_{1}(f_{20}-
 r_{2})}}                    
\end{equation}
As a numerical example let $r_{1}=r_{2}=-0.01$ and $f_{10}=f_{20}=0.02$. Then $\breve{N}_{1} /
 \hat{N_{1}} =0.6$, i.e. the number of followers of the ideology 1 descreases
 by 40\%. Of course this   causes some tension between the ideologies.  
 A measure of this tension can be through the index
 \begin{equation}\label{strain_meas1}
T_{i;k} = 1 - \frac{N_{i}^{(k)}}{\hat{N}_{i}},
\end{equation}
where $N_{i}^{(k)}$ is the population of the followers of the $i$-th ideology when
the $k$-th ideology is presented in the country too.
If the ideology is alone then $N_{1}^{(1)} = \hat{N}_{1}$ and the tension index is $T_{1;1}=0$.
If  $N_{1}$ decreases because of the competition with the second ideology,  then
the tension between the ideologies  characterised by  the tension index $T_{i;k}$ increases. 
The above definition for the tension holds even if $N_{1}$ follows some time dependent
trajectory. 
\par
The tension index can be generalized for the case of an  arbitrary number of ideologies
in the country (next section for example). Let $m$ ideologies be presented in the country. 
The tension on $i$-th ideology in presence of two other 
ideologies, $k$ and $l$ is
\begin{equation}\label{strain_meas2}
T_{i;k,l} (t) = 1 - \frac{N_{i}^{(k,l)}(t)}{\hat{N}_{i}}
\end{equation}
where $N_{i}^{(k,l)}$ is the population of followers of the $i$-th ideology when the
ideologies $k$ and $l$ operate in the country too.
The tension on $i$-th ideology in presence of three other ideologies, $j$, $k$, $l$ is
\begin{equation}\label{strain_meas3}
T_{i;j,k,l} (t) = 1 - \frac{N_{i}^{(j,k,l)}(t)}{\hat{N}_{i}}
\end{equation}
where $N_{i}^{(j,k,l)}$ is the number of the followers of the $i$-th ideology
in presence of ideologies $j$, $k$, $l$. In such a way we can define a series of
indices for  the quantification of the tensions among   ideologies competing for
followers ( in the same country). 
\par
Fig. 2 shows typical results from the numerical investigation of (\ref{mod1}) -(\ref{mod4}).
Fig. 2a shows the  purely inertial growth of a population of  followers of  
ideology 2 and its decline, followed by the inertial growth of the population of followers of
ideology 1. In Fig. 2b one can observe an initial decline  followed by an  inertial growth of 
the number of followers of  $both$ ideologies. 
These results can be usefully compared to a model of telecommunication competition 
\cite{legara}  where it is concluded that  (we quote) "schemes targeting local cliques within 
the network are more successful at gaining a larger share of the population than those that 
target users randomly at a global scale (e.g., television commercials, print ads, etc.). 
This suggests that success in the competition is dependent not only on the number of 
individuals in the population but also on how they are connected in the network.
The network in our above investigation is such that all agents are fully connected with each 
other; we consider a fully connected graph. This is equivalent to a mean field approximation 
study. Notice that the links are weighted through the $f_{ij}$ and $b_{ijk}$ coefficients.
\begin{figure}
\centering
\includegraphics[height=5cm,width=5cm]{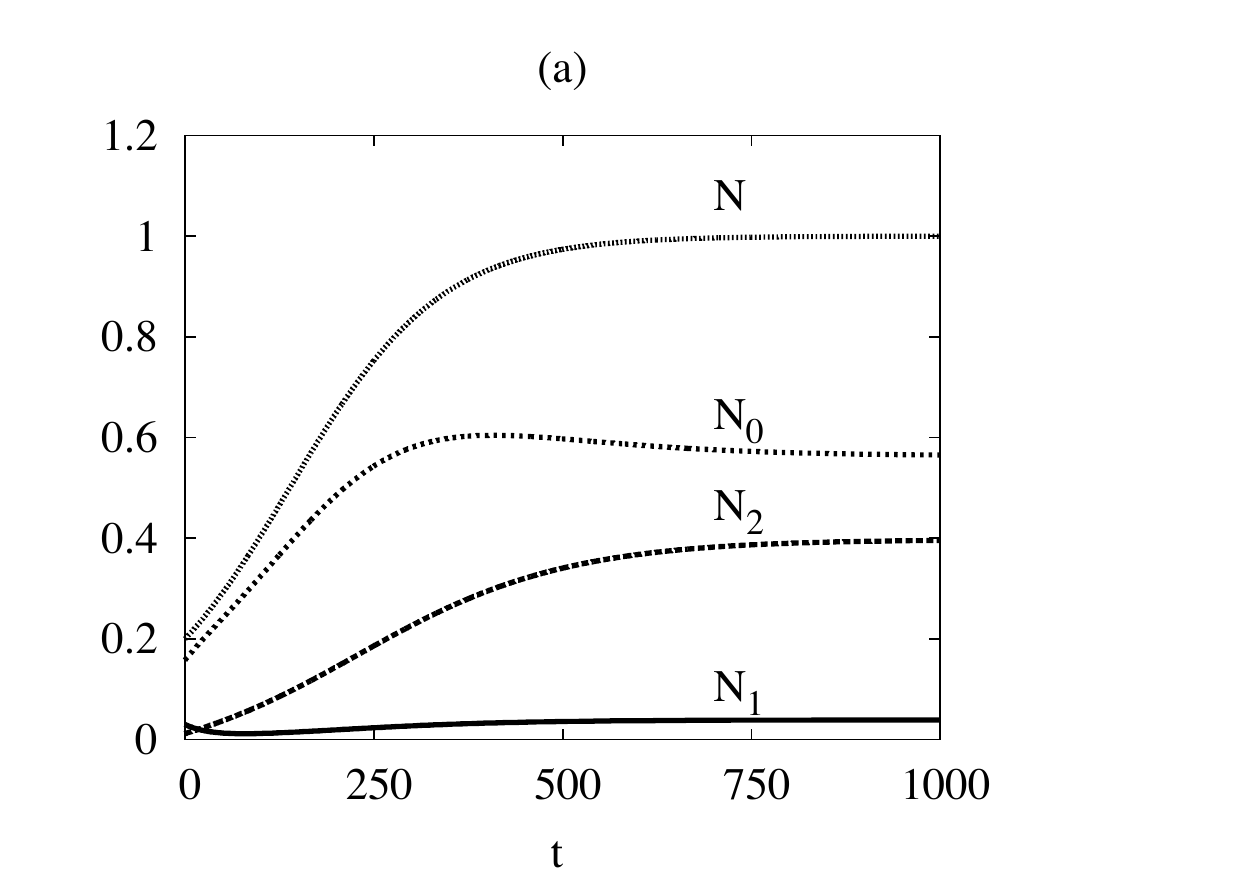}
\includegraphics[height=5cm,width=5cm]{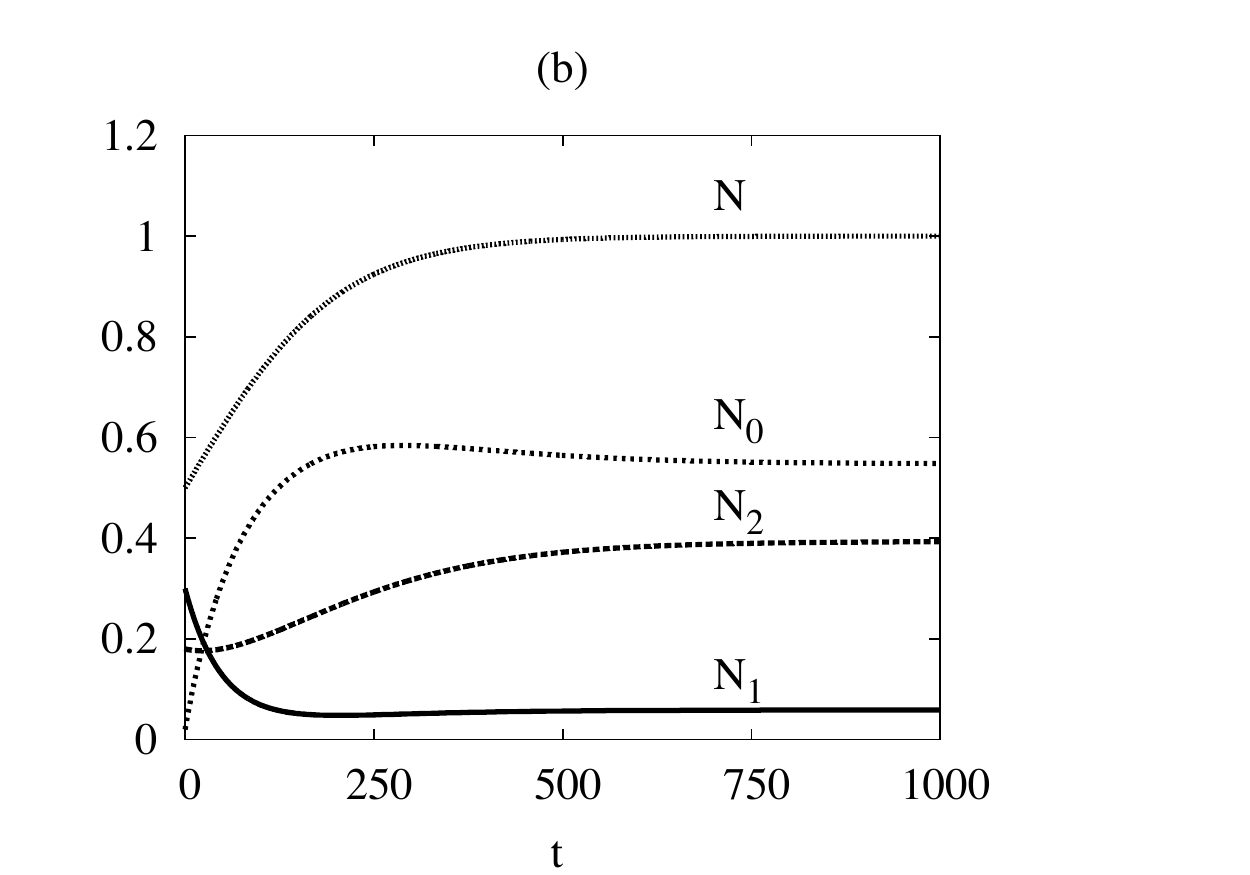}
\caption{\label{Fig7}     Evolution of the  total population $N$, number of followers of the ideologies, 
$N_{1}$ and $N_{2}$   and the number of  $N_{0}$ of people that are not followers of any 
ideology.  (a):
$r_{1}=-0.03$. Initial conditions: 
$N_{1}(0)=0.03$, $N_{2}(0)=0.012$, $N(0)=0.2$;  (b): $r_{1}=-0.02$.
Initial conditions:
$N_{1}(0)=0.3$, $N_{2}(0)=0.18$, $N(0)=0.5$.
In both figures :   $r_{2}=-0.005$, $r_{3}=0.01$, $C=1$, $f_{10}=0.001$, $f_{20}=0.003$,
$f_{11}=f_{22}=0$, $f_{12}=f_{21}=0.001$, $b_{10}=b_{20}= 0.001$, $b_{111}=b_{222}=0$;
$b_{112}=b_{121}=b_{122}=b_{211}=b_{212}=b_{221}=0.001$.}
\end{figure}
\section{Case of three ideologies}
The case of three ideologies will not be treated in full here. It can be easily understood 
that the above procedure can be extended to the case of multiple ideologies. We only restrict 
our presentation to the main  points and illustrate the newness.  In the case of three 
ideologies the model system (\ref{mod_syst1}) - (\ref{mod_syst3}) 
has an analytical solution  in 
the following conditions. First, let the binary conversion be negligible: $b_{ijk}=0$,
$b_{i0}=0$. Then the model system reduces to
\begin{equation}\label{analyt1}
N = N_{0} + \sum_{i=1}^{n} N_{1}
\end{equation}
\begin{equation}\label{analyt2}
\frac{d N}{d t} = r N \left( 1 - \frac{N}{C} \right)
\end{equation}
\begin{equation}\label{analyt3}
\frac{d N_{i}}{d t} = (r_{i} - f_{i0}) N_{i} + f_{i0} N + \sum_{j=1;j \ne i}^{n}
 (f_{ij}-f_{i0}) N_{j}
\end{equation}
Let the initial conditions  be $N (t=0) = N(0)$, $N_{i}(t=0) = N_{i}(0)$ and
$N(0) >C/2$. Then the solution is
\begin{equation} \label{asol1}
N(t) = \frac{C N(0)}{N(0) + (C-N(0)) e^{-rt}}
\end{equation}
\begin{eqnarray}\label{asol2}
N_{i}(t) = e^{-(f_{i0}-r_{1})t} \bigg \{ N_{1}(0) + \nonumber\\
\frac{c f_{i0}}{r} \bigg [ 
{\Phi} \bigg ( -\frac{C-N(0)}{N(0)},1,-\frac{f_{i0}-r_{i}}{r} \bigg ) -
\nonumber \\
e^{t(f_{i0}-r_{i})} {\Phi} \bigg(-\frac{C-N(0)}{N(0) e^{rt}},1,
- \frac{f_{i0}-r_{i}}{r}\bigg) \bigg] \bigg\}
\end{eqnarray}
\begin{equation}\label{asol3}
N_{0}(t)=N(t)- \sum_{i=1}^{n} N_{i}(t)
\end{equation}
where $\Phi$ is the special function defined in section 3. Above
$i=1,2,\dots,n$. When $n=1$ one returns to the case of section 3. For $n=2$
one recovers the solution of   the case of section 4.

\begin{figure}
\centering
\includegraphics[height=5cm,width=5cm]{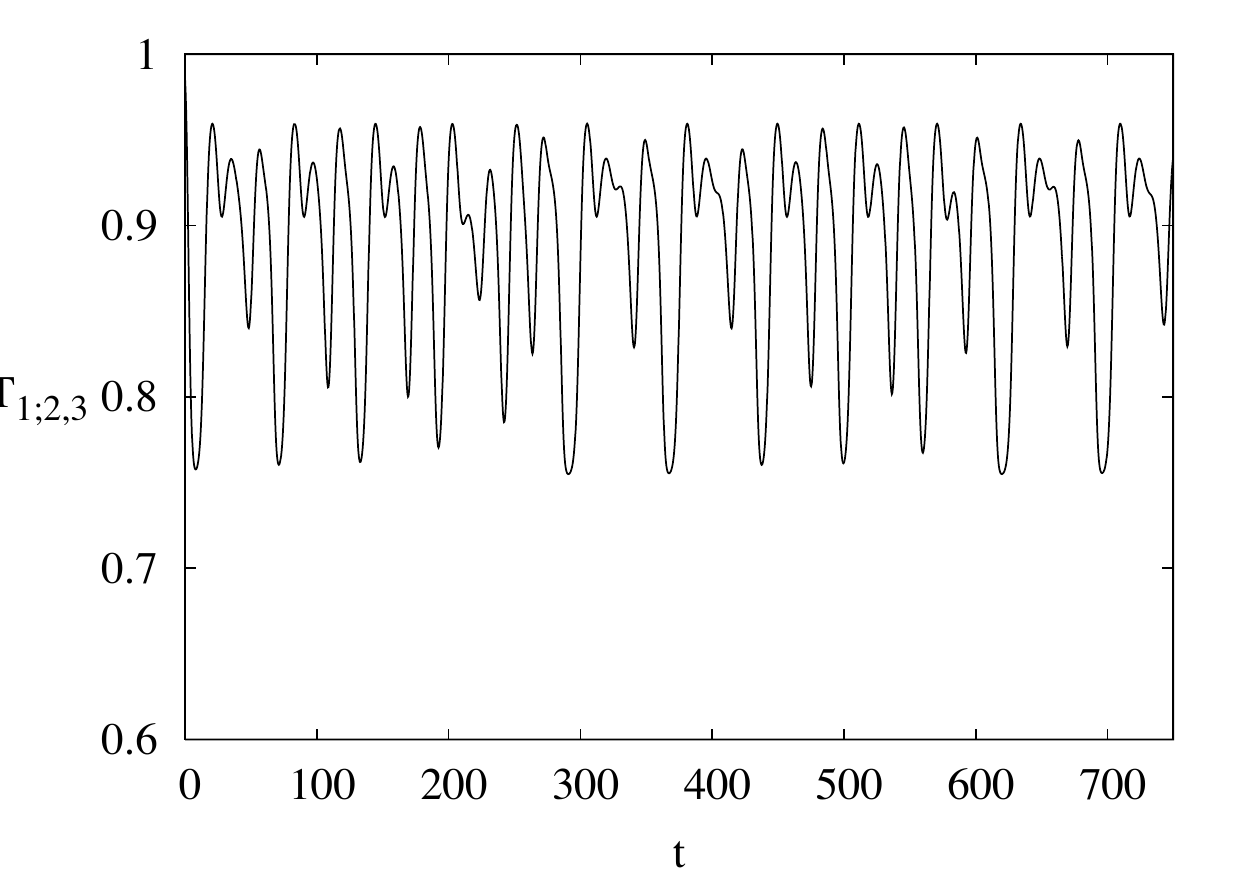}
\includegraphics[height=5cm,width=5cm]{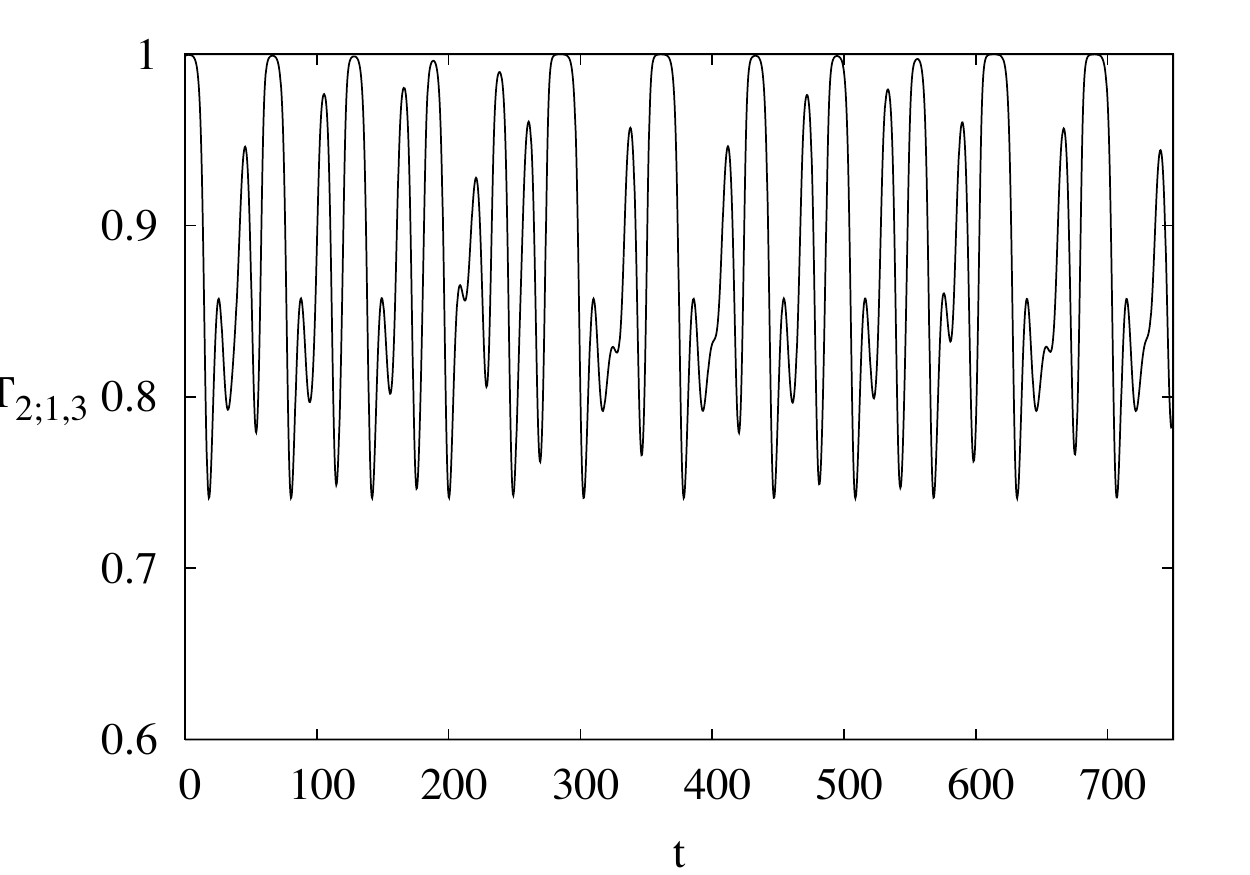}
\includegraphics[height=5cm,width=5cm]{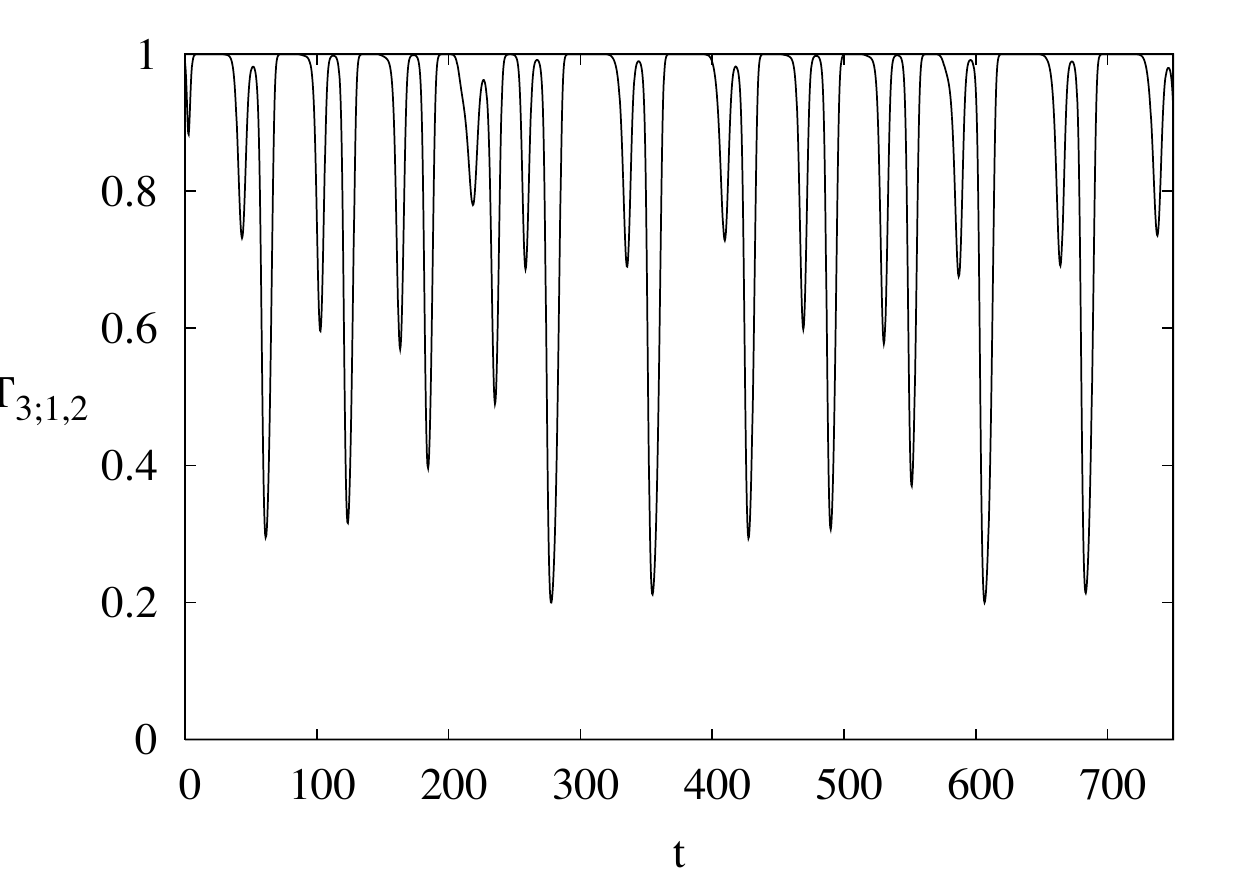}
\caption{\label{Fig6} An example  of chaotic behavior of the   tension indices  for the case 
of three competing ideologies. 
The model system is as in Eq.(\ref{gen_arneodo}). For the attractor shown the parameters
are as follows: $\kappa_{11}=\kappa_{1}$,$\kappa_{12}=\kappa_{1}$,
$\kappa_{13}=\kappa_{2}$, $\kappa_{21}=-\kappa_{1}$,
$\kappa_{22}=-\kappa_{2}$, $\kappa_{23}=\kappa_{2}$,
$\kappa_{31}=\kappa_{3}$, $\kappa_{32}=\kappa_{2}$,
$\kappa_{33}=\kappa_{2}$. $C=10$, $\kappa_{1}=0.5$, $\kappa_{2}=0.1$, $\kappa_{3}=1.63$,
$b_{10}=0.001$, $b_{20}=0.002$, $b_{30}=0.001$, $r_{1}=-0.01$,$r_{2}=-0.03$,$r_{3}=-0.01$.}
\end{figure}  

\par
Chaotic change of the ideological tensions 
becomes possible when the number of the ideologies becomes
equal to $3$ or larger. 
This can be demonstrated easily as follows.
Let the total population having  reached the carrying capacity of the
environment $N=C$. Let  $f_{12}=f_{13}=f_{10}=0$ and $b_{1jk}=b_{2jk}=b_{3jk}=0$
for $j,k=1,2,3$. Let also $f_{21}=f_{23}=f_{20}=0$ and 
$f_{31}=f_{32}=f_{30}=0$.  Let us rewrite the parameters in the following way
$$r_{1}+f_{11}= \kappa_{11}+\kappa_{12}+\kappa_{13},
b_{111}-b_{10}=-\kappa_{11}$$
$$ b_{112}-b_{10}=-\kappa_{12},
b_{113}-b_{10}=-\kappa_{13}$$
$$
r_{2}+f_{22} = \kappa_{21}+ \kappa_{22}+\kappa_{23},
b_{222}-b_{20}=-\kappa_{22}
$$
$$
b_{221}-b_{20}=-\kappa_{21}, b_{223}-b_{30}=-\kappa_{23}
$$
$$
r_{3}+f_{33} = \kappa_{31}+ \kappa_{32}+\kappa_{33},
b_{331}-b_{30}= - \kappa_{31}
$$
$$
b_{332}-b_{30}=-\kappa_{32}, b_{333}-b_{30}=-\kappa_{33}
$$
The model system of equations
becomes
\begin{equation}\label{gen_arneodo}
\frac{d N_{i}}{d t} = N_{i} \sum_{j=1}^{3} \kappa_{ij} (1-N_{j}) + b_{i0} C N_{i} ;
\hskip.5cm i,j=1,2,3. 
\end{equation} 
The existence of chaos in a particular case of (\ref{gen_arneodo}) was discussed
in \cite{arneodo}. Fig. 3 is an illustration of a case of chaotic change of the 
ideological tensions for the   (\ref{gen_arneodo}) corresponding model. 
$T_{1;2,3}$ is always significantly different from $0$.
This means that because of  the competition with the ideologies $2$ and $3$ the ideology
$1$ remains at a  significant "distance" from its most favorable state, - this one which 
would exist  if which there was no competitor. In addition
the ideology 1 copes relatively good with the situation as its tension
index remains relatively distant from $1$. The other two ideologies experience large tensions
and from time to time are close to extinction. The ideology 2 evolves better than the
ideology 3 which experiences large oscillations of the number of followers as
consequence of the competition with the other ideologies. This illustration for a 
given set of parameters indicates the interest of the approach,
since a few measures would allow to calibrate the parameters, in specific situations, whence would lead to considerations pertaining to forecasting science.
\section{Concluding remarks}
In this paper a general model for   ideological competition was formulated.
The model applies to cases in which countries have a  variable total population
which evolves according to the generalized Verhulst model. 
The discussion in the paper is concentrated on the  cases of
 constant coefficients and  when the total population of the country
increases.  An original ingredient concerns also the number of  followers of an ideology which can increase without
interpersonal contacts, but solely on the basis of so called
unitary conversion, e.g.  as a result of different forms of mass communication.  
The number of followers can also increase  by means of  binary conversion as
a result of interpersonal contacts. It is emphasized that the conversion can be outside the  
competing ideologies of interacting agents.
\par
The dynamics of the populations of  followers of the ideologies is discussed
for the case of one ideology and for the case of two and three competing ideologies, in Sect.  3-5, respectively.
 For the case of one ideology, the  
simple version of the general model
describes the evolution to an equilibrium state in which the population 
consists of some amount of  followers of the ideology and
persons indifferent to the ideology. The introduction of a second ideology
leads to some tension between the ideologies as the numbers of  followers drop
in comparison to the case when each of the ideologies is alone in the country.
The  ideological tensions can be quantified by a set of indices. 
A nonzero index   is a characteristic feature of the competition. Each ideology most of the 
time, if not always, tries to set its indices
of tension to $0$, i.e.
it tries to reach its maximum number of followers (which is the case when the
ideology  is alone in the country). This can be done by decreasing the number of  
followers of the other ideology on the territory. 
\par
We have  indicated that chaos can exist  (can be found) when the number of   available ideologies
increases above 2. The number of parameters is considerable, as in many realistic population 
evolution studies. However  the set of parameters appears to be realistic enough to be 
calibrated  in specific situations. This would lead to forecasting considerations.
This and the other obtained results hint to good perspectives 
for applications of the methods of statistical physics, theory of networks, 
sociophysics, etc \cite{jmiskdelay06,jmiskdelay07,alipomemory} to the 
problems of ideological competition.  This will be a subject of future research.

                \acknowledgments
We would like to thank the ESF Action COST MP0801 (Physics of Competition and Conflict)
for support of our research. Special thanks are devoted to  Noemi Olivera, Andrew Roach, 
J{\"u}rgen Mimkes and John Hayward for   stimulating discussions
on the dynamics of  ideological competition.

\end{document}